\documentclass[epj]{svjour}
\usepackage{graphics}
\usepackage{graphicx,ifpdf}
\usepackage{amsmath,amssymb}
\usepackage[numbers,sort&compress]{natbib}

\begin{document}
\title{3D reconstruction of dynamic liquid film shape by optical grid deflection method}
\titlerunning{3D reconstruction of dynamic liquid film shape by optical grid deflection method}

\author{L. Fourgeaud\inst{1,2}\thanks{Present address: Mechanical Design Office, Airbus Defence and Space, 31~rue des Cosmonautes, 31402 Toulouse Cedex 4, France}, E. Ercolani\inst{2}, J. Duplat\inst{2}, P. Gully\inst{2} \and V.~S. Nikolayev\inst{3}}
\authorrunning{L. Fourgeaud \textit{et al.}}

\institute{
  \inst{1} PSA, Route de Gisy, 78140 V\'elizy-Villacoublay, France\\
  \inst{2} Universit\'e Grenoble Alpes, CEA, INAC, Service des Basses Temp\'eratures, 38000 Grenoble, France\\
  \inst{3} Service de Physique de l'\'Etat Condens\'e, CEA, CNRS, Universit\'e Paris--Saclay, CEA Saclay, 91191 Gif-sur-Yvette Cedex, France
}

\date{Received:  / Revised version: }
%
\abstract{In this paper, we describe the optical grid deflection method used to reconstruct 3D profile of liquid films deposited by a receding liquid meniscus. This technique uses the refractive properties of the film surface and is suitable for liquid thickness from several microns to millimeter. This method works well for strong interface slopes and changing in time film shape; it applies when the substrate and fluid media are transparent. The refraction is assumed to be locally unidirectional. The method is particularly appropriate to follow the evolution of parameters such as dynamic contact angle, triple liquid-gas-solid contact line velocity or dewetting ridge thickness.
%
} 

\maketitle
\section*{Introduction}

Dynamics of dewetting films, spreading drops and moving triple liquid-gas-solid contact lines can be assessed with various experimental methods. Most of those are optical techniques which are suitable for studying interface movement because they are non-intrusive and do not disturb the fluid. We present here a method we used to accurately measure the temporal evolution of the topography of evaporating liquid films.

The idea is to obtain local properties of a heterogeneous medium by measuring the visible displacement of the object points created by the light refraction in the medium. To our knowledge, this principle has been first proposed by Kurata \textit{et al.} \cite{Kurata90} to measure the deformation of a water interface and by Gurfein \textit{et al.} \cite{Gurfein91} to characterize spatial variations of the refraction index of a thin transparent fluid layer. In \cite{Gurfein91}, the \emph{parallel} light rays crossed first the fluid layer and were refracted by it. Then the light fell to a grid of parallel wires situated close to this cell so that both were in focus of a camera behind. The grid image displacement in the direction perpendicular to it was proportional to the gradient of the fluid refraction index in the same direction. A similar setup has been used by Hegseth \textit{et al.} \cite{John} who have adapted it to observe the liquid film shape in a space-based experiment. Incident light rays were refracted at the film surface and the grid image distortion gave information about the film thickness. However the precision was weak for several reasons; in particular, the grid was out of focus.

In a setups used by Kurata \textit{et al.} and Andrieu \textit{et al.} \cite{Andrieu95}, a square grid pattern was drawn with an ink at the bottom of a thick transparent solid substrate. The latter was backlit with a \emph{diffuse} light and the liquid droplet is placed at the top. The camera was situated above the droplet, far away from it. The light rays were refracted at the liquid air interface and the grid image was distorted. The authors suggested measuring the grid distortion very close to the contact line to obtain the contact angle in dynamics. As one will see below, this method lacks precision when the slopes become large because one cannot measure the displacement at the contact line (only at some distance from it). In the works of Banaha \textit{et al.} \cite{Banaha09} and Kajiya \textit{et al.} \cite{Kajiya11}, a grid projection method has been used to track the contact line position of a water droplet on gels and to measure the droplet profile.

More recently, the grid was suggested to be replaced by a randomly placed points pattern \cite{Moisy09}. This method works well for small surface slopes and does not require any supplementary information (unlike the grid method as we will explain later) on the direction of the local film slope. The method however does not expect to work at large slopes (like those in the contact line vicinity) where the points would need to be placed densely so the correspondence between the object and image points is difficult to be established.

Snoeijer \textit{et al.} \cite{Snoeijer06a} have used a single wire to reconstruct the 2D film profile. Unlike other experiments, they have placed the light source and the camera at the same side of the fluid interface and used the wire reflection in the mirror polished substrate surface so the light ray is refracted twice by the film surface.

The main objective of the described setup is the 3D shape reconstruction of liquid films with contact lines. Such a reconstruction allows to determine precisely the local apparent contact angles, which is necessary to understand the contact line motion (i.e. drying) dynamics.

\section*{Fluid cell}

\begin{figure}
	\centering
	\includegraphics[width=0.4\textwidth, clip]{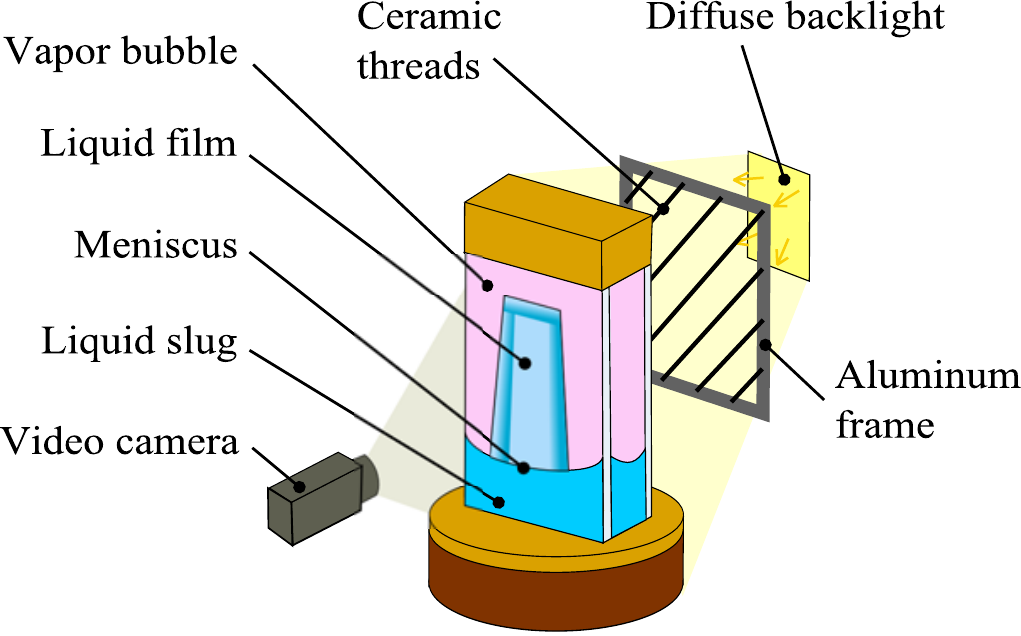}
	\caption{General scheme of the optical grid deflection technique setup.}
	\label{sch:principle}
\end{figure}

Fig.~\ref{sch:principle} presents the optical setup that is most appropriate for our case. The light source put behind the fluid cell is diffuse. The grid is placed between the source and the cell. The distance $d_g$ between the liquid film and the grid controls the grid deflection and for this reason needs to be adjustable. A video camera is set up far from the cell so that the lens collects the parallel light rays. A square patterned grid \cite{Andrieu95} is not required; the grid with inclined parallel threads is used to reduce the image treatment complexity. The grid inclination (45$^\circ$ with respect to the vertical axis) is chosen in such a way that it is never parallel to the contact line.

\begin{figure}[!ht]
	\centering
	\includegraphics[width=\columnwidth, clip]{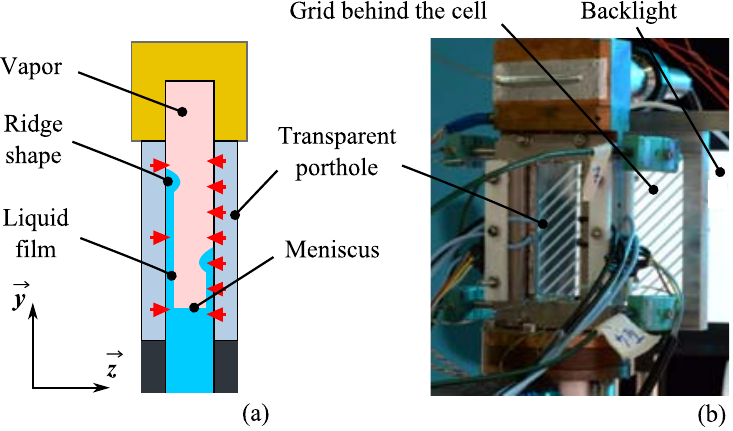}
	\caption{(a) Scheme of the experimental cell cross-section and (b) its photo.}
	\label{fig:cell}
\end{figure}

The fluid cell is a vertical closed Hele-Shaw cell made of two parallel sapphire portholes separated by gap of the spacing $d_c=2$~mm (Fig. \ref{fig:cell}). Inside the cell, a liquid meniscus of ethanol oscillates in its pure vapor atmosphere. During its receding, the meniscus deposits a liquid film on both portholes. These portholes are heated thanks to ITO (indium-tin oxide) transparent resistive layers. The back porthole is heated stronger than the other so the film deposited on it evaporates quicker and the front film (camera side) can be investigated alone.

The grid is formed by ceramic (alumina 99.9~$\%$) threads of rectangular cross-section of 2~mm thickness in the image (i.e. $xy$) plane. They are straight because of their rigidity and have regular and sharp edges. The threads are equally spaced due to the high precision machining of the aluminum frame. As the distortion near the contact line is strong, the number of threads is limited in order to prevent an overlap between the distorted and non-distorted threads. The light source is a white light LED panel (50000~cd/m$^2$). The video camera is equipped with a CMOS $2/3^{\prime\prime}$ sensor of resolution 2048$\times$1088. As the film oscillation frequency is 1.5 Hz, the camera is used with the frame rate of 280 images per second.

\section*{2D film deformation}

\begin{figure}
	\centering
	\includegraphics[width=0.45\textwidth, clip]{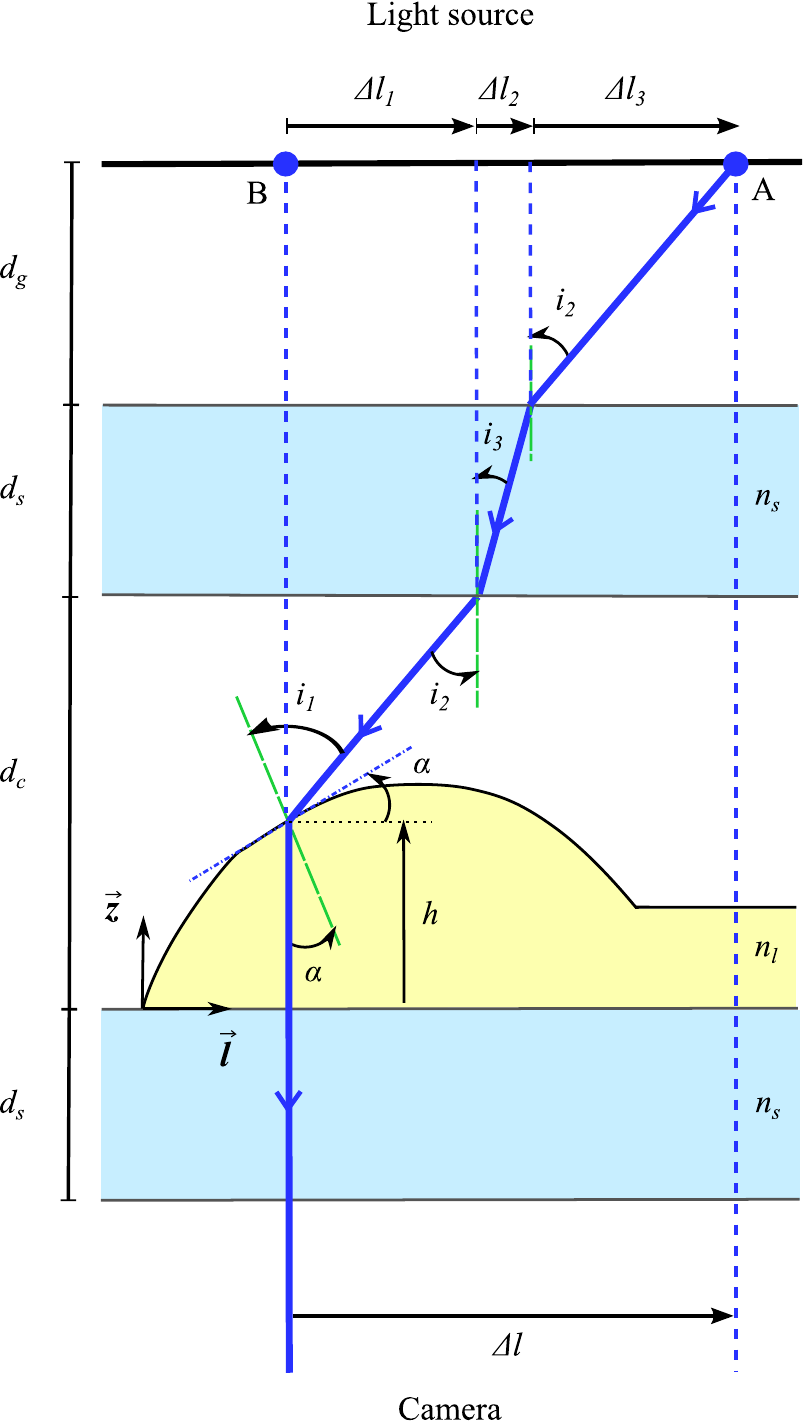}
	\caption{Ray tracing in the plane ($\vec{l}$,$\vec{z}$) orthogonal to the triple contact line and to the porthole plane. In reality, $d_g$ is much larger than both $d_s$ and $d_c$, and $d_c = d_s$.}
	\label{sch:trajet_lum}
\end{figure}

To explain the film shape reconstruction, let us assume first that the film is deformed along a single direction $\vec{l}$, i.e. that its deformation is 2D (Fig.~\ref{sch:trajet_lum}). An incident light ray is refracted by several interfaces. Point A is the object point at the grid thread; point B is its image viewed by the video camera. $\alpha$ is the angle of the tangent to the film and $h=h(l)$ is its local thickness. $d_g$ and $d_s$ represent the distance between the grid and the cell and the thickness of substrate, respectively.

The camera situates far from the cell and its lens is focused at infinity; we assume that the collected rays are parallel to the optical axis. As the porthole plane (i.e. the film substrate) is perpendicular to the optical path, the ray is not refracted at its interfaces. The last refraction occurs at the film interface as shown in Fig. \ref{sch:trajet_lum}.

The ray deflection from point A to point B is denoted $\Delta l = l_A - l_B$. One applies the refraction law at each interface. Angles are oriented positively in the trigonometric direction. The refractive indexes of the ethanol vapor and of the air are both equal to 1. In the liquid ethanol and in the sapphire substrate, refractive indexes are respectively $n_l=1.355$ and $n_s=1.78$. On the film free surface:
\begin{eqnarray}\label{eq:1}
\sin (i_1) & = &  n_l \sin (\alpha).
\end{eqnarray}
On the interface vapor/substrate and substrate/air:
\begin{eqnarray}\label{eq:2}
  \sin (i_2) & = & n_s \sin (i_3),
\end{eqnarray}
where $i_2=i_1 -\alpha$ (Fig.~\ref{sch:trajet_lum}). Deflections created in each medium read:
\begin{eqnarray}
		\Delta l_1 & = & (d_c - h)\tan (i_2) \\
		& = & \simeq d_c \tan(i_2)\\
		\Delta l_2 & = & d_s \tan (i_3) \\
		\Delta l_3 & = & d_g \tan (i_2)
\end{eqnarray}
The total deflection $\Delta l$ is a sum of three contributions,
\begin{eqnarray}
\Delta l & = & \Delta l_1 + \Delta l_2 + \Delta l_3
\end{eqnarray}
The last contribution is the largest, $\Delta l_3\gg \Delta l_1 + \Delta l_2$, because $d_g=6$~cm while $d_c = d_s=2$~mm. Eqs.~\eqref{eq:1} and \eqref{eq:2} serve to express $\Delta l$ as a function of the angle $\alpha$,
\begin{eqnarray}\label{eq:3}
\Delta l & \simeq & d_g \tan \left( \arcsin (n_l \sin \alpha ) - \alpha \right).
\end{eqnarray}
As $n_l > 1$, $\Delta l$ is of the same sign as $\alpha$. Knowing $\Delta l$, numerical solution of Eq.~\eqref{eq:3} gives the angle $\alpha$. Since $\alpha$ is defined in the Cartesian reference $(\vec{l}, \vec{z})$,
\begin{eqnarray}\label{eq:hContinu}
\dfrac{d h}{d l} & = & \tan \alpha \text{,}
\end{eqnarray}
where $l$ is the coordinate of the point B associated to the $\vec{l}$ axis. The integration of this last equation leads to $h$ as a function of $l$.

Fig.~\ref{sch:trajet_lum} shows that each point B of the thread image corresponds to a particular film point. The geometrical construction of Fig.~\ref{sch:trajet_lum} also shows that each point B corresponds to a unique point A of the thread. However the inverse statement is wrong. An object point A might correspond to different image points (which may exist if a ray with a different angle $i_2$ of incidence results in an outcoming ray parallel to the optical axis) or correspond to no image point at all.

\section*{3D reconstruction}
To reconstruct the film profile, one needs to determine a unique object point A corresponding to each point B of the grid image. This is however not trivial when the film thickness $h$ depends both on $x$ and $y$. In this case the direction $\vec l$ of Fig.~\ref{sch:trajet_lum} is that of the gradient of $h$ (i.e., the steepest slope direction that varies in space). However one cannot determine the direction $\vec l$ for each point B$_i$ from the image because the grid threads are continuous; one needs an additional information or hypothesis.

\begin{figure}
	\centering
	\includegraphics[width=0.45\textwidth, clip]{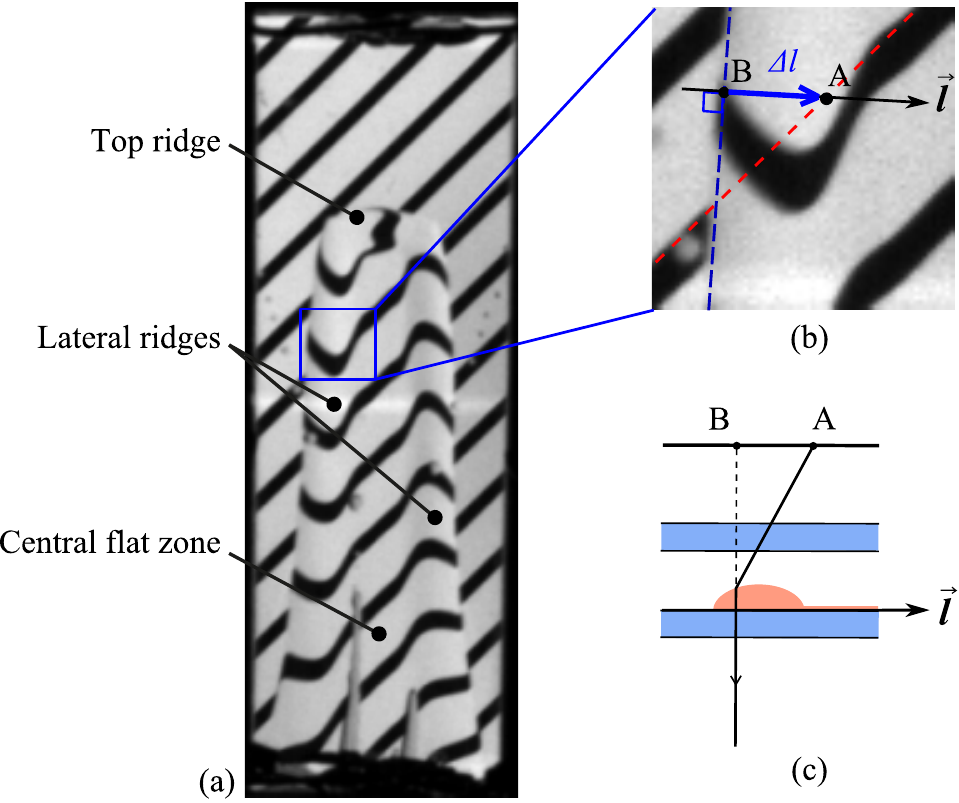}
	\caption{(a) Different zones of the film on the film image. (b) Zoom on the lateral ridge. (c) Ridge cross-section corrresponding to the image (b).}
	\label{sch:intro}
\end{figure}

Fig.~\ref{sch:intro}a presents a typical image of our cell. The trapezoidal film shape is common in the capillary dewetting of the flat surface \cite{Snoeijer06a,Gao15}. In the present case, the substrate is heated above the saturation temperature so the film evaporation occurs. The origin of the dewetting phenomenon observed at evaporation is explained in our preceding article \cite{PRF16}. While the overall evaporation is weak, the evaporation mass flux varies sharply over the film area. The evaporation rate is especially strong near the contact line \cite{Wayner,Raj2012}, which creates high apparent contact angles \cite{EuLet12} in spite of the complete wetting conditions at equilibrium. The high contact angle, in its turn, causes the capillary dewetting phenomenon and a ridge-like structure along the contact line (cf. Fig.\ref{fig:cell}a) similarly to the non-wetting situation with no evaporation \cite{Snoeijer06a}.

Because of the high thermal conductivity of sapphire, its temperature is nearly homogeneous. This means that the evaporation is nearly invariable along the contact line, which is thus straight with the ridge profile invariable along it. Indeed, one can see in Fig.~\ref{sch:intro}a that the ridge zone width variation along the contact line is small. In the ridge zone one can thus assume a 2D film profile in the direction orthogonal to the contact line (associated with the $\vec{l}$ axis) by neglecting the film slope in the direction tangential to the contact line. In other words, $h$ will be assumed to depend only on $|\vec{l}|$, and not on the coordinate on the direction tangential to the contact line. The validity of such an assumption will be evaluated below.

In Fig.~\ref{sch:intro}b, the red line represents the contour of the object thread. The blue arrow shows the deflection $\Delta l$ created along the $\vec{l}$ axis shown in black. The schematized cross section along the $\vec{l}$ axis is represented in Fig.~\ref{sch:intro}c.

The direction of the axis $\vec{l}$ may vary from one image point to another. In Fig.~\ref{sch:intro}, three zones where the contact line is straight can be distinguished: on the top ridge zone, the contact line is horizontal; on the lateral ridge zones, it is inclined with respect to $\vec{y}$ direction. Consequently, it is possible to define only three different $\vec{l}$ directions corresponding to each ridge zone (Fig.~\ref{sch:def_reperes}). Note that the thread images are not deflected in the central zone. This means that the liquid film does not refract the light there and thus is nearly flat \cite{LauraATE17}.

\section*{Definition of the axes}
\begin{figure}[!ht]
	\centering
	\includegraphics[width=0.45\textwidth, clip]{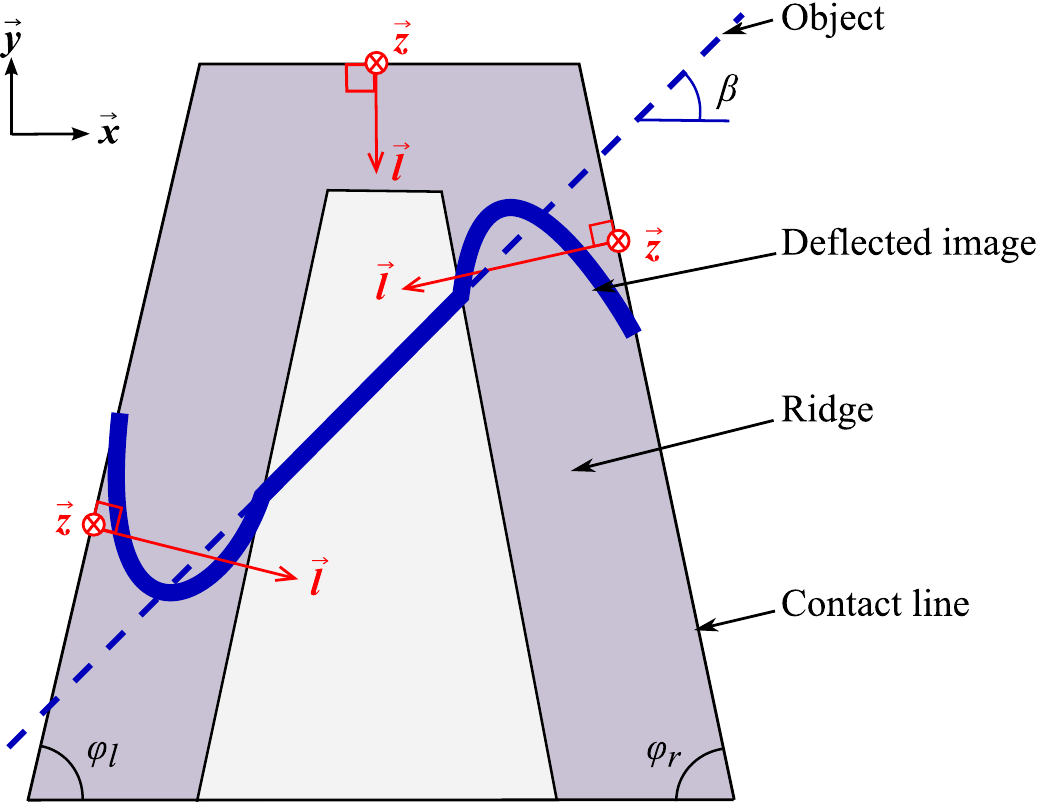}
	\caption{Definition of the axes to study each ridge zone.}
	\label{sch:def_reperes}
\end{figure}

From Fig.~\ref{sch:intro}, we define a Cartesian reference $(x,y)$. The $x$ axis is oriented from left to right and vector $\vec{y}$ from bottom to top (Fig.~\ref{sch:def_reperes}). The origin of this reference system situates at the image left-bottom corner. For each ridge zone, we define also a basis $(\vec{l}, \vec{z})$ belonging to the plane orthogonal to the contact line. Vector $\vec{l}$ belongs to the plane $(x,y)$. It is oriented inside the liquid film in agreement with Fig.~\ref{sch:trajet_lum}. Vector $\vec{z}$ is orthogonal to the plane $(\vec{x}, \vec{y})$ and is oriented from the outside to the inside of the cell (Fig.~\ref{sch:def_reperes}) so that $h>0$. It gives the direction of optical axis. Besides, both lateral parts of the ridge are inclined. Their contact lines are assumed to be straight (as discussed above) and form angles $\varphi_l$ and $\varphi_r$ with $\vec{x}$. In the following, these angles will be denoted $\varphi_j\geq 0$, $j$ can take the value ${l,r,t}$ to refer to the left, right or top ridge zone.

The direction $\vec{l}$ depends on the considered image point $\rm{B}_i$. The unit vector along it is defined as $\vec{e_l} = (\sin\varphi_l, -\cos\varphi_l)$ for the left side, $\vec{e_l} = (-\sin\varphi_r, -\cos\varphi_r)$ for the right side and $\vec{e}=(0,-1)$ for the top side.

\section*{Deflection calculation}
\begin{figure}[!ht]
	\centering
	\includegraphics[width=0.45\textwidth, clip]{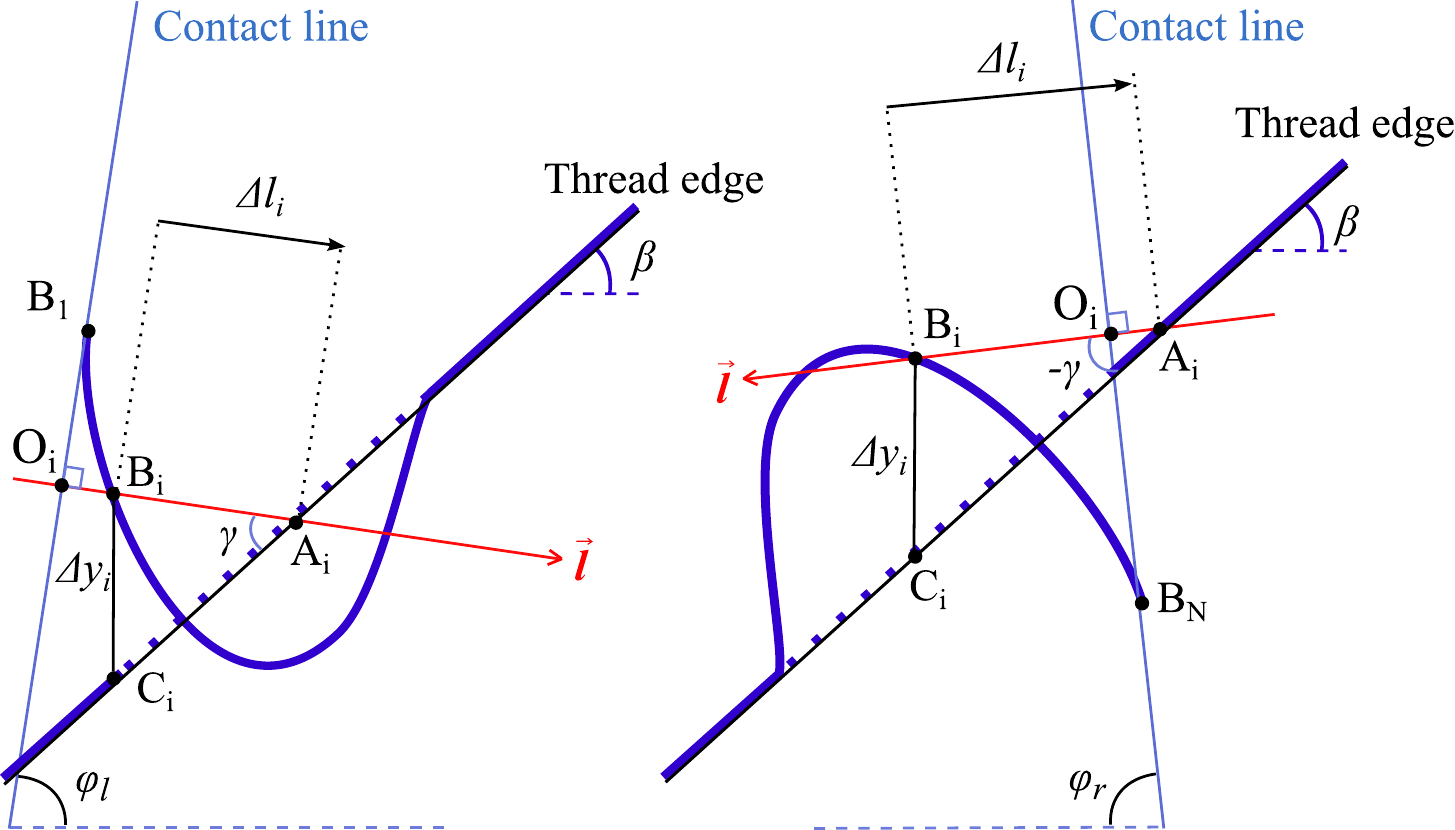}
	\caption{Geometrical scheme of deflection at the lateral left (cf. Fig.~\ref{sch:intro}b for the corresponding image) and right parts of the ridge, respectively.}
	\label{sch:DL}
\end{figure}

The set of Eqs~\ref{eq:3} and~\ref{eq:hContinu} is sufficient to obtain the 3D profile $h(l)$ once the displacement $\Delta l$ is known. Fig.~\ref{sch:DL} illustrates the geometrical construction to obtain $\Delta l_i$ for a point $\rm{B_i}$ which is the image of the point $\textrm{A}_i$. Point $\rm{C_i}$ is the vertical projection of point $\rm{B}_i$ onto the reference thread. The distance between points $\rm{B}_i$ and $\rm{C}_i$ is $\Delta y_i = y_B - y_C$.

We introduce $\beta_l = \beta$ and $\beta_r = - \beta$ according to the ridge part. In the following, it will be merely noted $\beta_j$ with $j=\lbrace l,r \rbrace $. Whatever the observed side, the sine theorem applied to the triangle ABC gives
\begin{eqnarray}\label{eq:4}
\dfrac{1}{\Delta l_i}\sin (\dfrac{\pi}{2}-\beta) & = & \dfrac{1}{\Delta y_i}\sin (\gamma)
\end{eqnarray}
We note that $\gamma = \dfrac{\pi}{2} - \varphi_j + \beta_j $. According to Eq.~\eqref{eq:4},
\begin{eqnarray}\label{eq:5}
\Delta l_i & = & \dfrac{\cos(\beta)}{\cos(\beta_j - \varphi_j)} \Delta y_i
\end{eqnarray}
Finally, by measuring the distance $\Delta y_i$, one deduces $\Delta l_i$ from Eq.~\eqref{eq:5}.
For the top ridge part,
\begin{eqnarray}\label{eq:5bis}
\Delta l_i & = & \Delta y_i.
\end{eqnarray}
$\Delta y_i$ can be measured on film pictures. Thanks to Eq.~\eqref{eq:5} and \eqref{eq:5bis}, $\Delta l_i$ is deduced from $\Delta y_i$. And $\Delta y_i$ is then used in Eq.~\eqref{eq:3} to obtain the $\alpha_i$ angle for each point of the thread, which is necessary to solve Eq.~\eqref{eq:hContinu}.

\section*{Thickness evolution}
The numerical integration of Eq.~\eqref{eq:hContinu} requires to know the $l_i = \textrm{O}_i\textrm{B}_i$ coordinate of each B$_i$ pixel on the thread image, where $i$ varies from 1 to N. Points B$_1$ and B$_\textrm{N}$ (Fig.~\ref{sch:DL}) belong to the left and right parts of the contact line, respectively. $\textrm{O}_i$ is the origin of the axis $\vec{l}$ defined for each B$_i$ point. $\textrm{O}_i$ is defined as the intersection of the axis $\vec{l}$ with the contact line. In the absolute reference $(\vec{x}, \vec{y})$, point B$_i$ coordinates are denoted $(x_i ; y_i)$, with $i = 1\ldots \textrm{N}$.

For the left ridge part,
\begin{eqnarray}
l_{i} & = & \overrightarrow{\rm{B}_1\rm{B}_i}.\vec{e_l} \nonumber \\
& = & (x_i - x_1)\sin\varphi_l - (y_i-y_1)\cos\varphi_l.
\end{eqnarray}
On the right side we obtain (by definition of points B$_1$ et B$_\textrm{N}$):
\begin{eqnarray}\label{eq:OBd}
l_{i} & = & \overrightarrow{\text{B}_\textrm{N}\text{B}_i}.\vec{e_l} \nonumber \\
& = & - (x_i - x_\textrm{N})\sin\varphi_r - (y_i-y_N)\cos\varphi_r.
\end{eqnarray}
On the top ridge part:
\begin{eqnarray}\label{eq:7}
l_{i} & = & y_\textrm{N} - y_i.
\end{eqnarray}

By knowing $l_i$ abscissa for each point of the distorted perimeter, one discretizes Eq.~\eqref{eq:hContinu}. Integration is performed from the film center (where the initial condition is given by the measured by interferometry film thickness $h_c$ as explained below) towards the edges. We note that $l_1=l_\textrm{N}=0$. On the left side:
\begin{eqnarray}\label{eq:hDiscretg}
h_{i-1} & = & (l_{i-1}-l_i)\dfrac{\tan \alpha_i + \tan \alpha_{i-1}}{2} + h_i.
\end{eqnarray}
On the right and top parts, the formula is different, because the points numbering is performed in the opposite to $\vec{l}$ direction:
\begin{eqnarray}\label{eq:hDiscretd}
h_{i+1} & = & -(l_{i+1}-l_i)\dfrac{\tan \alpha_i + \tan \alpha_{i+1}}{2} + h_i.
\end{eqnarray}
We note that the film thickness $h_i$ is determined at the points B$_i$ $(x_i, y_i)$ (i.e. along the distorted image of the thread) and not at the points A$_i$ (along the straight reference thread line). The film points $(x_i, y_i, h_i)$ are obtained for both the upper and the lower thread borders. Typical 3D profiles are presented below: Fig.~\ref{sch:raw_pictures} shows the evolution of the film at three time moments. The 3D reconstruction of each picture is given in Fig.~\ref{graph:3D}.

\textit{Infine}, every thickness points obtained are plotted on the same graph in 3D (blue dotted lines on Fig~\ref{graph:3D}). The 3D reconstruction is completed by using  interpolation (colored meshes).
\begin{figure}[!ht]
	\centering
	\includegraphics[width=0.46\textwidth, clip]{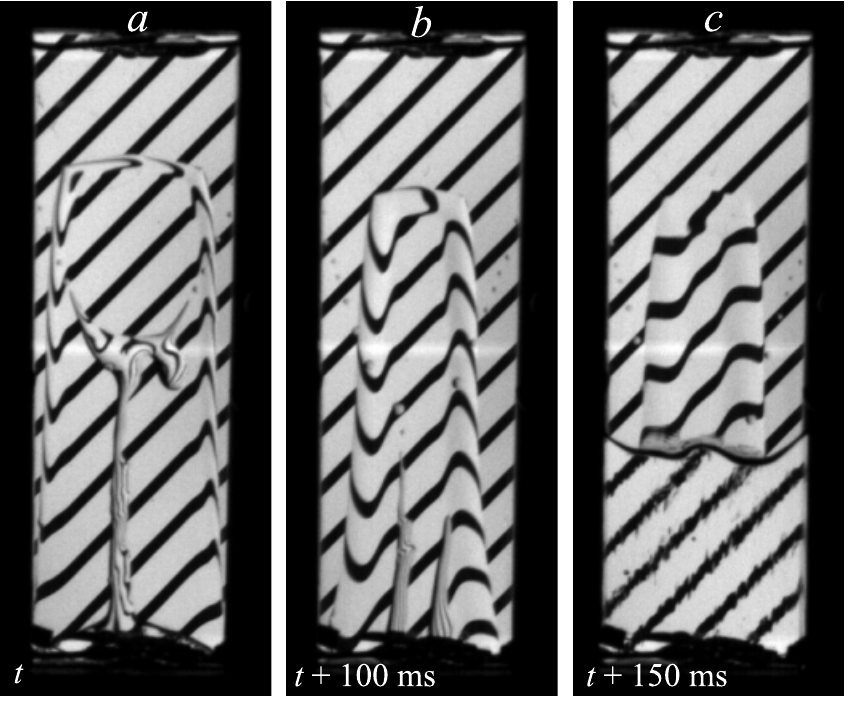}
	\caption{Film images at different time moments.}
	\label{sch:raw_pictures}
\end{figure}
\begin{figure}[!ht]
	\centering
	\includegraphics[width=0.4\textwidth, clip]{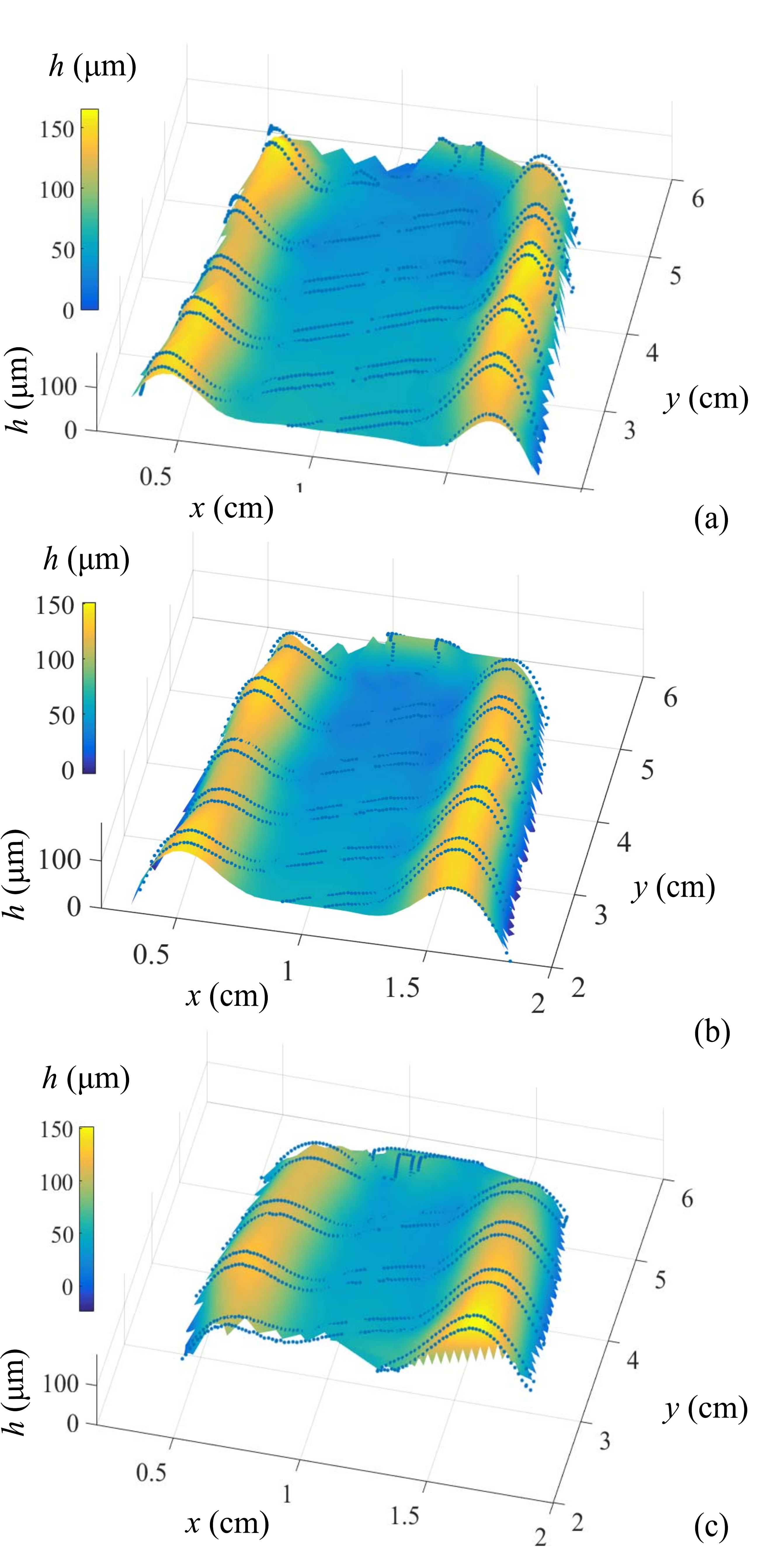}
	\caption{3D film shape reconstruction from the respective images shown in Fig~\ref{sch:raw_pictures}.}
	\label{graph:3D}
\end{figure}

\section*{Thickness of the central film part}
The central film part is flat. In fact the film slope exists, but is very small, of the order of $0.6^\circ$ \cite{LauraATE17}. Because of this, its thickness $h_c$ can be measured by the Ocean Optics HR2000+ spectrometer with a probe that integrates both white light emitting and light sensor parts. If illuminated normally to its surface, the film forms a kind of Fabry-Perot interferometer. The rays that are reflected at the cell-liquid and liquid-vapor interfaces interfere and the resulting spectrum is analyzed by the device. The spectrum maxima occur when the phase difference
\begin{equation}\label{dp}
\Delta \phi\equiv\frac{2 \pi}{\lambda_m}p  =  2 \pi m
\end{equation}
with
\begin{equation}\label{eq:h_0}
p=2 n_l h_c,
\end{equation}
the optical path difference, $m$, an arbitrary integer number, and $\lambda_m$, the corresponding wavelength of the spectrum maximum. From Eq.~\eqref{dp}, one obtains $\lambda_m^{-1}=m/p$. By plotting the inverse wavelengths of the spectrum maxima versus their consecutive numbers, one thus obtains a straight line. The film thickness can be inferred from the slope with Eq.~\eqref{eq:h_0}. This leads to the film thickness value just in front of the probe. 
As the interferometer time response is small enough (10~ms) with respect to the film oscillation period (0.5~s), the thickness is known in real time. This measurement is synchronized with the image acquisition, and all the data are combined to achieve the final reconstruction. The measurement uncertainty has been evaluated to be $\pm~3~\rm{\mu m}$ by using different samples of known thickness.

\section*{Apparent contact angle}

\begin{figure}[!ht]
	\centering
	\includegraphics[width=0.3\textwidth, clip]{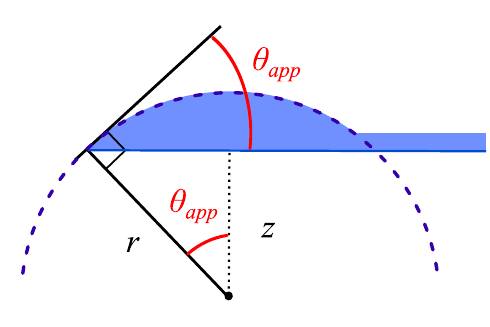}
	\caption{Geometrical construction used to deduce the apparent contact angle $\theta_{app}$. The flat film area and the ridge are represented in blue; the osculating circle is dotted.}
	\label{sch:osc}
\end{figure}

It is difficult to reconstruct the film profile in the contact line vicinity and an extrapolation is needed to obtain the contact angle with a sufficient precision. From previous studies \cite{deG,Snoeijer06a,Snoeijer08a}, we know that the ridge should be a cylindrical segment; our data confirm that it is so (Fig.~\ref{sch:osc}). It is thus possible to fit the ridge reconstruction with a circle. The extrapolation of circle allows us determining the contact angle $\theta_{app}$ between the ridge and the substrate as shown in Fig.~\ref{sch:osc},
\begin{equation}
\theta_{app}  = \arccos \dfrac{z}{r},
\end{equation}
with $z$ the height of the circle center with respect to the substrate surface and $r$ the circle radius.
\begin{figure}[!ht]
	\centering
	\includegraphics[width=0.4\textwidth, clip]{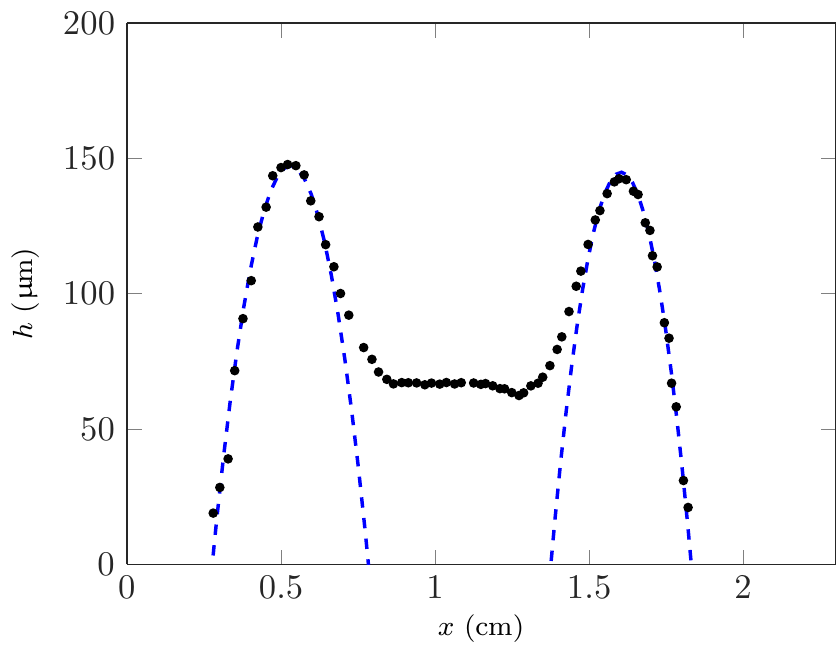}
	\caption{Film cross section at the height $y = 3$~cm from Fig.~\ref{graph:3D} along $\vec{x}$. Dotted lines: left and right parts of the ridge fitted with a circle. Note the scale difference between $x$ and $y$-axis that causes the circle deformation. For this case, $\theta_{app}= 9^{\circ}$.}
	\label{graph:osc}
\end{figure}

\section*{Error bar}
The optical grid deflection method has been validated thanks to prisms of known in advance angles. The influence of refractive indexes (of liquid ethanol and sapphire) and the accuracy of grid contour detection have been estimated: the global uncertainty on the $\alpha$ angle is $\pm~2^\circ$.

However, the dominant error source seems to be the hypothesis about the absence of the film slope in the direction tangential to the contact line. In reality, the ridge is somewhat larger at the bottom compared to the top. This can be seen in Fig. \ref{sch:raw_pictures}. The related error can be evaluated as follows. The resulting ridge thickness uncertainty is noted $\delta h$ and is proportional to the ridge thickness variation $\delta L$, let us say on a scale of the total height $\delta y$ of the film. From  Eq. \eqref{eq:hContinu},
\begin{equation}
\delta h  =  \delta L \tan (\alpha)
\end{equation}
where $\alpha$ is the value averaged over the ridge. $\delta L$ is related to $\delta y$,
\begin{equation}
\delta L  =  \delta y \frac{L_b - L_t}{y_t-y_b},
\end{equation}
where $L_b$ is the ridge width at the film bottom $y_b$ and $L_t$ is the ridge width at the film top $y_t$. Over the total film height, the global error does not exceed $\pm~10$\%.

\section*{Conclusion}
The grid deflection method described here is based on the light refraction at the strongly curved liquid film surface. It is possible to provide the 3D reconstruction of a curved interface provided that the direction of maximum slope is known, which is the case of the liquid ridge observed e.g. in dewetting liquid films. The method is non-intrusive. The main advantages of this technique are simplicity and good accuracy. The dynamic reconstruction is possible due to the good temporal resolution of the method. The reconstruction can be performed by using a sequence of pictures in order to measure the simultaneous evolution of parameters such as dynamic contact angle and ridge profile.

\section*{Acknowledgments}
We thank A. Gauthier for his technical assistance throughout this project. We are grateful to B. Andreotti for his advice concerning the film thickness measurements and to D. Garcia and F. Bancel for their help. We thank V. Padilla for machining the grid frame. The financial contributions of ANR within the project AARDECO ANR-12-VPTT-005-02, of CNES within the ``Fundamental and applied microgravity'' program and of ESA within MAP INWIP are acknowledged.

\section*{Author contribution statement}
All authors contributed to this study. LF, EE, PG contributed to the development of experimental installation. LF, JD and EE conducted the experiment. LF, VN and JD developed the theory of 3D reconstruction and analyzed the results of experiments. LF and VN have prepared the manuscript, with input from all other authors.


\end{document}